\documentclass[fleqn,usenatbib]{mnras}

\usepackage{newtxtext,newtxmath}
\usepackage[T1]{fontenc}

\DeclareRobustCommand{\VAN}[3]{#2}
\let\VANthebibliography\thebibliography
\def\thebibliography{\DeclareRobustCommand{\VAN}[3]{##3}\VANthebibliography}

\usepackage{graphicx}	% Including figure files
\usepackage{amsmath}	% Advanced maths commands
\title[The Fast UV Transient AT2020xnd]{Real-time Discovery of AT2020xnd:  A Fast, Luminous Ultraviolet Transient with Minimal Radioactive Ejecta}

\newcommand{\ljmu}{1}
\newcommand{\berkeley}{2}
\newcommand{\miller}{3}
\newcommand{\caltech}{4}
\newcommand{\eso}{5}
\newcommand{\stockholm}{6}
\newcommand{\iitbo}{7}
\newcommand{\lsstc}{8}
\newcommand{\iitbe}{9}
\newcommand{\uw}{10}
\newcommand{\icecsic}{11}
\newcommand{\warsaw}{12}
\newcommand{\turku}{13}
\newcommand{\tuorla}{14}
\newcommand{\cardiff}{15}
\newcommand{\ipac}{16}
\newcommand{\birmingham}{17}
\newcommand{\dtuspace}{18}
\newcommand{\southampton}{19}
\author[D. A. Perley et al.]{
Daniel A. Perley,$^{\ljmu}$\thanks{E-mail: d.a.perley@ljmu.ac.uk}
Anna Y. Q. Ho,$^{\berkeley,\miller}$
Yuhan Yao,$^{\caltech}$
Christoffer Fremling,$^{\caltech}$
\newauthor
Joseph P. Anderson,$^{\eso}$
Steve Schulze,$^{\stockholm}$
Harsh Kumar,$^{\iitbo,\lsstc}$
G. C. Anupama,$^{\iitbe}$
\newauthor
Sudhanshu Barway,$^{\iitbe}$
{Eric C. Bellm},$^{\uw}$
Varun Bhalerao,$^{\iitbo}$
Ting-Wan Chen,$^{\stockholm}$
\newauthor
Dmitry A. Duev,$^{\caltech}$
Llu\'is Galbany,$^{\icecsic}$
Matthew J. Graham,$^{\caltech}$
Mariusz Gromadzki,$^{\warsaw}$
\newauthor
Claudia P. Guti\'errez,$^{\turku,\tuorla}$
Nada Ihanec,$^{\warsaw}$
Cosimo Inserra,$^{\cardiff}$
Mansi M. Kasliwal,$^{\caltech}$
\newauthor
Erik C. Kool,$^{\stockholm}$
S. R. Kulkarni,$^{\caltech}$
Russ R. Laher,$^{\ipac}$
Frank J. Masci,$^{\ipac}$
James D. Neill,$^{\caltech}$
\newauthor
Matt Nicholl,$^{\birmingham}$
Miika Pursiainen,$^{\dtuspace}$
Joannes van Roestel,$^{\caltech}$
Yashvi Sharma,$^{\caltech}$
\newauthor
Jesper Sollerman,$^{\stockholm}$
Richard Walters,$^{\caltech}$
and
Philip Wiseman$^{\southampton}$
\\
$^{\ljmu}$ Astrophysics Research Institute, Liverpool John Moores University, IC2, Liverpool Science Park, 146 Brownlow Hill, Liverpool L3 5RF, UK\\
$^{\berkeley}$ Department of Astronomy, University of California, Berkeley, 94720, USA\\
$^{\miller}$ Miller Institute for Basic Research in Science, 468 Donner Lab, Berkeley, CA 94720, USA\\
$^{\caltech}$ Division of Physics, Mathematics, and Astronomy, California Institute of Technology, Pasadena, CA 91125, USA\\
$^{\eso}$ European Southern Observatory, Alonso de Córdova 3107, Casilla 19 Santiago, Chile\\
$^{\stockholm}$ Department of Astronomy, The Oskar Klein Centre, Stockholm University, AlbaNova, 10691, Stockholm, Sweden\\
$^{\iitbo}$ Indian Institute of Technology Bombay, Powai, Mumbai 400076, India \\
$^{\lsstc}$ LSSTC DSFP Fellow \\
$^{\iitbe}$ Indian Institute of Astrophysics, II Block Koramangala, Bengaluru 560034, India \\
$^{\uw}$ {DIRAC Institute, Department of Astronomy, University of Washington, 3910 15th Avenue NE, Seattle, WA 98195, USA} \\
$^{\icecsic}$ Institute of Space Sciences (ICE, CSIC), Campus UAB, Carrer de Can Magrans, s/n, E-08193 Barcelona, Spain. \\
$^{\warsaw}$ Astronomical Observatory, University of Warsaw, Al. Ujazdowskie 4, 00-478 Warsaw, Poland \\
$^{\turku}$  Finnish Centre for Astronomy with ESO (FINCA), FI-20014 University of Turku, Finland \\
$^{\tuorla}$  Tuorla Observatory, Department of Physics and Astronomy, FI-20014 University of Turku, Finland \\
$^{\cardiff}$ School of Physics \& Astronomy, Cardiff University, Queens Buildings, The Parade, Cardiff CF24 3AA, UK\\
$^{\ipac}$ IPAC, California Institute of Technology, 1200 E. California Blvd, Pasadena, CA 91125, USA  \\
$^{\birmingham}$ Birmingham Institute for Gravitational Wave Astronomy and School of Physics and Astronomy, University of Birmingham, Birmingham B15 2TT, UK \\
$^{\dtuspace}$ DTU Space, National Space Institute, Technical University of Denmark, Elektrovej 327, 2800 Kgs. Lyngby, Denmark \\
$^{\southampton}$ School of Physics and Astronomy, University of Southampton, Southampton SO17 1BJ, UK \\
}

\date{Accepted 2021 September 21. Received 2021 September 21; in original form 2021 February 17}

\pubyear{2021}

\begin{document}
\label{firstpage}
\pagerange{\pageref{firstpage}--\pageref{lastpage}}
\maketitle

\begin{abstract}
The many unusual properties of the enigmatic AT2018cow suggested that at least some subset of the empirical class of fast blue optical transients (FBOTs) represents a genuinely new astrophysical phenomenon.  Unfortunately, the intrinsic rarity and fleeting nature of these events have made it difficult to identify additional examples early enough to acquire the observations necessary to constrain theoretical models.  We present here the Zwicky Transient Facility discovery of AT2020xnd (ZTF20acigmel, the ``Camel'') at $z=0.243$, the first unambiguous AT2018cow analog to be found and confirmed in real time.   AT2018cow and AT2020xnd share all key observational properties: a fast optical rise, sustained high photospheric temperature, absence of a second peak attributable to ejection of a radioactively-heated stellar envelope, extremely luminous radio, millimetre, and X-ray emission, and a dwarf-galaxy host.  This supports the argument that AT2018cow-like events represent a distinct phenomenon from slower-evolving radio-quiet supernovae, likely requiring a different progenitor or a different central engine.  The sample properties of the four known members of this class to date disfavour tidal disruption models but are consistent with the alternative model of an accretion powered jet following the direct collapse of a massive star to a black hole.  Contextual filtering of alert streams combined with rapid photometric verification using multi-band imaging provides an efficient way to identify future members of this class, even at high redshift.
\end{abstract}


%%%%%%%%%%%%%%%%%%%%%%%%%%%%%%%%%%%%%%%%%%%%%%%%%%

%%%%%%%%%%%%%%%%% BODY OF PAPER %%%%%%%%%%%%%%%%%%
\clearpage

\section{Introduction}

A typical supernova rises on a timescale of days to weeks and fades away on a timescale of weeks to months \citep{Villar+2017,Perley+2020}.  For stars that explode as supergiants, the rise timescale is governed by the cooling of the shock-heated photosphere and the fading timescale is governed by the gradual recombination of the ejecta \citep{Arnett1980,Weiler2003,Zampieri2017}.  For stars that explode in a compact state (stripped-envelope Wolf-Rayet stars and white dwarfs), the emission from the shock breakout and cooling is primarily at X-ray wavelengths and the rise and fall in the optical band are instead dominated by the dispersal of heat from newly-synthesized radioactive elements through the expanding ejecta \citep{Arnett1982}.

Over the past decade, a population of transients with fast rise times ($t_{\rm rise} \sim 1-7$ days), fast decay times ($t_{\rm decline,1/2} \sim 3-12$ days), and a range of peak optical luminosities ($-16 \gtrsim M_{g,{\rm peak}} \gtrsim -22$) has been uncovered by wide-area surveys, with the largest samples originating from Pan-STARRS (PS1; \citealt{Drout+2014}) and the Dark Energy Survey (DES; \citealt{Pursiainen+2018}).  These are sometimes called ``Rapidly Evolving Transients'' (RETs), ``Fast-Blue Optical Transients'' (FBOTs), or ``Fast-Evolving Luminous Transients'' (FELTs; \citealt{Rest+2018}).  Events with these properties simultaneously require an energetic shock, a large pre-explosion radius, and a low ejecta mass \citep{Inserra2019}.  While this combination of parameters is unusual it is not without precedent:  type IIb supernovae, which are thought to originate from the explosion of a compact star with an extended but tenuous hydrogen atmosphere, show an initial early shock-breakout peak similar in nature to FBOTs (e.g. \citealt{Fremling2019}).  However, the initial peak in type IIb SNe is not as luminous and it is followed by a second radioactively-powered peak of comparable optical luminosity that is not seen in FBOTs.  The poorly-understood class of Type Ibn supernovae \citep{Pastorello+2007} also shows many similarities to the DES/PS1 FBOTs \citep{Fox+2019,Xiang2021}.

The seminal event in the understanding of this class was the discovery of AT2018cow at 60 Mpc \citep{Prentice+2018}.  The rise to peak was very fast ($\lesssim$3 days from explosion to peak), it was extremely luminous at peak, and it faded quickly---properties characteristic of the PS1 and DES FBOTs.  However, after peak it displayed a number of unexpected and indeed unprecedented behaviours:  (a) the spectrum remained continuum-dominated throughout, with a high blackbody temperature ($>$10000\,K) and a photosphere that expanded rapidly before peak but then propagated inward \citep{Perley+2019};  (b) it was extremely luminous at radio and millimetre wavelengths, with a millimetre light curve that did not reach maximum for several weeks \citep{Ho+2019}, but displayed no significant millimetre polarisation or VLBI proper motion \citep{Huang+2019,Mohan+2020,Bietenholz+2020}; (c) it was also luminous at X-ray wavelengths, and showed an erratic light curve that flickered repeatedly up and down by a factor of ten in flux on timescales of days \citep{Ho+2019,Margutti+2019,Kuin+2019,Rivera+2018}.  Late-time spectra were dominated by intermediate-width ($\sim$4000 km~s$^{-1}$) lines of hydrogen and helium, showing some similarities with SNe Ibn \citep{Fox+2019}.

These properties impose several additional stringent requirements on the progenitor.  The lack of a second peak implies that it did not produce a large amount of radioactive nickel or unbound ejecta \citep{Perley+2019}.  The high radio luminosity and late millimetre peak imply a fast but nonrelativistic shock traversing dense circumstellar material (CSM) beyond the optical photospheric radius \citep{Ho+2019}. The rapid X-ray variability requires a compact and long-lived central engine (or perhaps a complex shock in a confined structure) that is either exposed to the viewer or lightly screened \citep{Margutti+2019,Ho+2019}.

A variety of models have appeared in the literature attempting to explain this combination of properties.  Electron-capture supernovae and fallback supernovae are commonly appealed to since both naturally explain the low ejecta mass, with either a proto-magnetar or an accretion-powered jet invoked to explain the fast rise and luminous X-ray/radio emission \citep{Perley+2019,Margutti+2019,Quataert+2019,Lyutikov+2019,Piro+2020,Uno+2020a}.  However, many other models exist \citep[e.g.,][]{Soker+2019,Mohan+2020,Leung+2020,Leung+2021}, including a range of models that associate AT2018cow with an unusual tidal disruption event (involving an intermediate-mass or even stellar-mass black hole) rather than an unusual supernova \citep{Kuin+2019,Perley+2019,Liu+2018,Uno+2020b,Kremer+2021}.

A challenge in distinguishing different models is the fact that only a single well-observed event exists (AT2018cow itself).  While the data set for this event is excellent, it is unknown whether any of its qualitative or quantitative properties are essential to the phenomenon (as opposed to peculiar features of this event alone).  It is plausible to assume that some of the DES and PS1 FBOTs represent the same generic phenomenon, but some members of these samples may be physically unrelated: for example, many exhibit a much lower peak luminosity or show evidence of cooling towards standard recombination temperatures, and no firm constraints exist on their behaviour outside the optical band.  Thus, while the DES/PS1 samples are quite large and have been analyzed in some detail \citep{Wiseman+2020}, the bulk sample properties cannot confidently be held to be indicative of the nature of AT2018cow.   As a result there is no firm constraint on the cosmic rate, typical host-galaxy environment, or degree of internal diversity among other examples of this phenomenon.

Recently, two additional AT2018cow-like objects have been reported in the literature: CSS161010 (\citealt{Coppejans+2020}) and ZTF18abvkwla (``Koala'', \citealt{Ho+2020}).  Both of these events show very luminous radio emission lasting for months following the optical event; both also originated from dwarf galaxies (the host of ZTF18abvkwla is very strongly star-forming, that of CSS161010 much less so).  Unfortunately, in neither case was the nature of the transient recognized early enough in its evolution to motivate a fast and deep optical campaign to establish the temperature evolution in detail\footnote{Some optical follow-up of CSS161010 was acquired but the light curve is not yet published.} or an early X-ray campaign to search for rapid variability.

In this paper, we present the discovery of AT2020xnd (ZTF20acigmel, a.k.a. ``The Camel''), a fast optical transient very similar to AT2018cow that was identified in real time using the Zwicky Transient Facility (ZTF; \citealt{Bellm2019,Graham2019}) via our custom search pipeline.  We outline our discovery process and present the deep optical follow-up observations that our early discovery enabled.  We demonstrate that, like AT2018cow, this event showed no classical, radioactively-powered supernova, it remained very blue until late times, and it was extremely luminous across the electromagnetic spectrum.  These properties suggest that the key features of AT2018cow are shared by other members of the class and indeed are likely the defining aspects of the phenomenon, separating AT2018cow and its ilk from other fast-rising transients.  Our discovery also provides a road-map for gradually building up samples of well-observed events in the coming years using high-cadence surveys.

\section{Observations}

\begin{figure}
	\includegraphics[width=8.5cm]{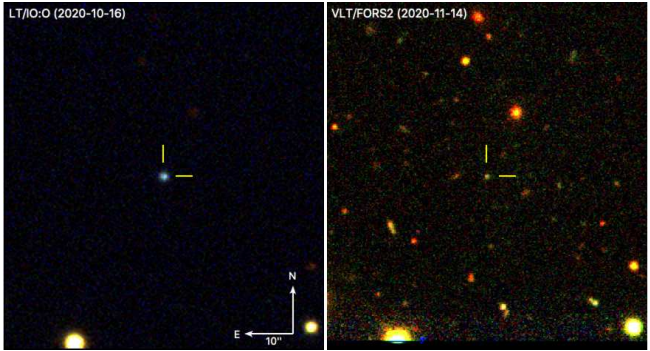}
    \caption{RGB false-colour $u$/$g$/$r$ image of the transient from the Liverpool Telescope 3 days after peak (left panel), compared to late-time VLT $u$/$g$/$R$ imaging 26 days after peak (right panel).   The source is distinctly blue at early times.  The host galaxy  dominates the flux in the VLT image.}
    \label{fig:image}
\end{figure}

\subsection{ZTF Discovery}

ZTF is an optical time-domain facility conducting a series of transient surveys \citep{Bellm2019surveys} using the 48-inch Oschin Schimdt Telescope at Palomar Observatory.  These include a public survey in $g$ and $r$ (conducted at three-day cadence through September 2020 and two-day cadence from October 2020) and a one-day cadence Caltech survey in $g$ and $r$.
The ZTF observing system is described in \citet{Dekany2020},
and images are processed with the ZTF real-time reduction and image subtraction pipeline at the Infrared Processing \& Analysis Center \citep{Masci2019}.
Each 5-$\sigma$ detection in a subtracted image is registered as an ``alert'' \citep{Patterson2019},
and each alert receives machine-learning based real-bogus scores, one based on a random-forest classifier \citep{Mahabal2019} and one based on a neural network \citep{Duev2019}.
The three nearest Pan-STARRS \citep{Flewelling+2020} sources receive a star-galaxy score to assist with identifying stellar vs. extragalactic transients \citep{Tachibana2018}.
The alert stream is distributed to a variety of community brokers; the one used for this paper was \texttt{kowalski}\footnote{https://github.com/dmitryduev/kowalski}.

One of the primary goals of ZTF, particularly with the higher-cadence surveys, is to find fast extragalactic transients such as gamma-ray burst afterglows and AT2018cow-like events.
To this end, a filter has been set up with the following criteria:

\begin{itemize}
    \item A deep-learning based real-bogus score exceeding 0.65
    \item At least two detections, with a duration between them exceeding 20 minutes
    \item A Galactic latitude exceeding 15\,deg
    \item A criterion to remove artifacts from nearby bright stars (similar to that employed in \citealt{Perley+2020})
    \item No coincident stellar counterpart (does not have a PS1 catalog match with a star-galaxy score exceeding 0.76 within 2\,arcsec)
    \item Detected at a magnitude brighter than 20\,mag
\end{itemize}

The resulting candidates are then sorted into four groups:

\renewcommand{\labelenumi}{\arabic{enumi}.}
\begin{enumerate}
    \item New transients: those with no previous detections prior to the current night.
    \item High-redshift transients: those with a DESI Legacy Imaging Survey DR8 (hereafter ``Legacy Survey'';  \citealt{Dey+2019}) counterpart within three arcseconds with a photometric redshift exceeding 0.4).
    \item Fast-peaking transients: transients with a light curve that has peaked, i.e. has pre- and post-detections 1-sigma below peak, and where the time from half-max to max is under five days, as per \citet{Ho+2020}.
    \item Fast-evolving transients: transients that rise more rapidly than 1 mag/day or fade more rapidly than 0.3 mag/day (see \citealt{Andreoni2020}).
\end{enumerate}

Every day, one of us (DAP, AYQH, YY) scans the resulting data stream, which usually has roughly 20 candidates.
In addition, once per week forced photometry is run on all transients from the previous week, to identify candidates missed because of sub-threshold detections.

ZTF20acigmel passed the filter on 2020-10-12 under criterion 2 above: it is a new source coincident with a faint extended Legacy Survey source (type ``REX'') with a high photometric redshift ($z=1.33^{+0.76}_{-0.40}$; \citealt{Zhou+2021}).
It was not immediately identified as a transient of interest during scanning: the event was relatively faint ($g\sim19.7$ mag, $r\sim20.1$ mag), and the most recent upper limit was three days prior and relatively shallow ($r>20.2$ mag). 
It passed the filter again on 2020-10-14 following a slight rise in flux and was also not saved.
On 2020-10-16 it passed the filter a third time, and by this time it has faded significantly from the peak, suggesting fast evolution.  (Both public-survey and Caltech one-day cadence survey data were aquired and contributed to all three of these epochs.) It was saved as a candidate and registered to TNS, and the Liverpool Telescope \citep{Steele2004} was triggered for follow-up observations.
Follow-up observations were organized and coordinated using the GROWTH Marshal \citep{Kasliwal2019}.

After the transient had faded, we re-ran forced photometry for all observations of the field using the average position for all measurements (J2000 coordinates $\alpha$=22:20:02.014, $\delta$=$-$02:50:25.35).  This photometry is given in Table~\ref{tab:photometry}.

\begin{figure*}
	\includegraphics[width=17.5cm]{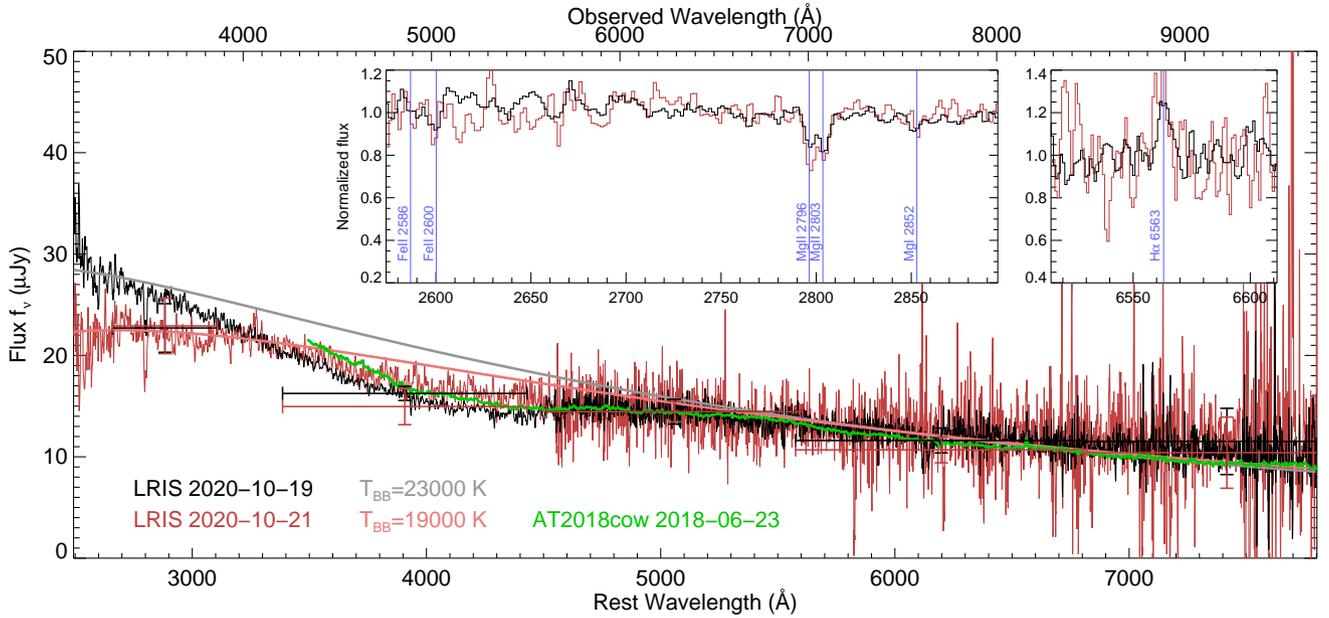}
    \caption{LRIS spectra of AT2020xnd, acquired two days apart and approximately one week after the peak of the optical light curve.  The first spectrum is shown in black; the second spectrum is shown in red and has been rescaled in flux to match the first (in $r$-band).  (Photometry from LT is overplotted as measurements with error bars.)  Theoretical blackbody spectra are also shown as thick curves, and a spectrum of AT2018cow at similar phase also plotted for comparison.  No unambiguous SN features are evident in either spectrum, although there is a hint of an extremely broad absorption feature centred around 4000\AA\ (rest frame) in the first spectrum.  The Mg II $\lambda\lambda$2796,2803 narrow absorption doublet is seen in both spectra along with possible low-significance detection of absorption lines of Mg I and Fe II (large inset), and weak emission from H$\alpha$ (small inset).}
    \label{fig:spectrum}
\end{figure*}

\subsection{LT Observations}

We obtained imaging observations using the 2m robotic Liverpool Telescope (LT) on two successive nights (beginning at approximately 2020-10-17 00:07 UT and 2020-10-17 20:34 UT, respectively), using 60-second exposures in all five Sloan Digital Sky Survey (SDSS) filters ($ugriz$).    AT2020xnd was well-detected in all bands in both of these observations (Figure~\ref{fig:image}).  (A longer observation was also obtained on 2020-10-20, although weather conditions were poor and only unconstraining upper limits were obtained.)  The magnitude of the transient was measured in all images using basic aperture photometry (aperture radius 1.5$^{\prime\prime}$), calibrated against SDSS secondary reference stars.  

Comparison to the P48 observations confirmed a rapid drop in flux: by $0.5\pm0.1$ mag over the 3 days between the epoch of apparent maximum light and the first LT observation, then another $0.3\pm0.1$ during just the next 20.5 hours (both measurements are in $g$-band).  The observed colour was very blue ($u-g=-0.42\pm0.05$ mag in the first LT epoch and $u-g=-0.52\pm0.10$ mag in the second LT epoch, after correcting for Galactic reddening of $E(B-V) = 0.07$ mag; \citealt{Schlafly+2011}).  Blue colours are commonly observed in young supernovae and in cataclysmic variables, but both the degree of the colour and its persistence almost 1 mag into the decline are unusual.  However, persistent blue colours after peak were a hallmark of AT2018cow \citep{Perley+2019}.  This, in combination with the Legacy Survey detection of a probable host galaxy, motivated additional follow-up, in particular spectroscopy (\S \ref{sec:lris}).

\subsection{P60 Observations}

Additional imaging observations were acquired with the Rainbow Camera (RC) of the Spectral Energy Distribution Machine (SEDM) on the Palomar 60-inch telescope \citep{Blagorodnova+2018}.  Observations were taken using all four filters ($ugri$) and reduced using the basic RC pipeline, and photometry was performed following the procedures of \cite{Fremling+2016}.  Due to the fading of the source and limited sensitivity of the detector only upper limits were obtained in $u$-band, but detections confirming continued fading of the source were secured in all three remaining filters on 2020-10-20 and 2020-10-21.  After that time, the event became too faint to secure useful detections with SEDM.

\subsection{GIT Observations}

On 2020-10-20 at 13:45 UT, we started imaging observations of AT2020xnd on the 0.7m GROWTH-India Telescope (GIT) located at the Indian Astronomical Observatory (IAO), Hanle-Ladakh (India). 
The data were acquired in $r^\prime$ band with multiple 300~sec exposures for four successive nights, although only an upper limit was obtained on 2020-10-23 due to poor observing conditions.  Reduction was performed using the standard GIT pipeline. 
PSF photometry of the transient was performed with PS1 stars as a reference.  

\subsection{NTT Observations}

Follow-up was provided using the ESO Faint Object Spectrograph and Camera (EFOSC2) on the 3.6m New Technology Telescope (NTT) at La Silla as part of the ePESSTO+ project \citep{Smartt+2015}.  Two epochs were acquired: the first on 2020-10-20 ($ugri$, under dark conditions) and the second on 2020-10-23 ($g$ and $r$ only, under bright conditions).  A basic reduction of the data was performed using IDL.  Photometry of AT2020xnd was performed using an aperture radius of 1.0 arcsec after subtraction of reference imaging from the VLT (\S \ref{sec:vlt}).

\subsection{VLT Imaging Observations}
\label{sec:vlt}

We obtained several epochs of imaging using FORS2 on the Very Large Telescope at Paranal, Chile, under DDT proposal 106.21U2.  Observations were obtained in the $G_{\rm high}$, $R_{\rm special}$, and $I_{\rm Bessel}$ filters on UT 2020-11-03, 2020-11-04, 2020-11-05, 2020-11-09, 2020-11-14, and 2020-12-06.  (Additionally, $U_{\rm high}$-band observations were obtained on 2020-11-14.)  An image of the field is shown in Figure~\ref{fig:image}.  Reference imaging of the host galaxy was acquired on 2021-05-10 ($R_{\rm special}$ and $I_{\rm Bessel}$) and 2021-06-07 ($G_{\rm high}$ and $U_{\rm high}$).  We reduce the data using the same procedures as for the NTT observations.

The host galaxy contributes most of the flux at the location of the transient at late times, and we employ a custom image subtraction routine to obtain uncontaminated measurements of the transient magnitude in all six VLT epochs and in all filters.  The FORS2 $U_{\rm high}$, $G_{\rm high}$, and $R_{\rm special}$, and $I_{\rm Bessel}$ filters are treated as $u_{\rm SDSS}$, $g_{\rm SDSS}$, $R_{c}$ and $I_{c}$ filters (respectively), with SDSS secondary standards transformed to the Cousins system using the Lupton relations \footnote{http://classic.sdss.org/dr4/algorithms/sdssUBVRITransform.html}.

The magnitudes of the host galaxy (measured in the VLT images using a 1.5 arcsec aperture, all in the AB system and not corrected for Galactic extinction) are $u=25.6\pm0.3$, $g=24.75\pm0.10$, $R=24.23\pm0.12$, $I=24.56\pm0.25$.
We correct the LT, GIT, and P60 photometry for host-galaxy contamination by subtracting the corresponding fluxes from the flux in the earlier imaging containing the SN.  (For consistency we also assume a host magnitude of 26 AB in the UVOT filters and subtract this from the UVOT fluxes.)   The resulting host-subtracted photometry is reported in Table~\ref{tab:photometry}.  This table excludes upper limits: a complete table of all photometry is available in machine-readable format as supplementary material.

\subsection{Keck/LRIS Spectroscopy}
\label{sec:lris}

We obtained spectroscopy of AT2020xnd using the Low-Resolution Imaging Spectrometer (LRIS; \citealt{Oke+1995}) on the Keck I Telescope on 2020-10-19 and 2020-10-21.  Observations were reduced using LPipe \citep{Perley2019}.  Both spectra (Figure~\ref{fig:spectrum}) show a very hot, blue continuum with no easily-identifiable broad supernova features.

A series of narrow absorption lines originating from the interstellar or intergalactic media are superimposed on the continuum of both spectra.  The strong Mg II $\lambda\lambda$2797,2801 doublet, redshifted to $z=0.2433$, is prominent in both observations.   Mg I $\lambda$2852 and Fe II $\lambda$2600 may also be present (at lower significance).  These features establish $z=0.2433$ as a minimum redshift.  No other absorption lines are observed, although a weak, narrow emission line of H$\alpha$ is also seen in both spectra at a consistent redshift, which if attributed to the host galaxy fixes this as the redshift of the transient.  We will assume $z=0.2433$ throughout this paper.  (We assume a basic cosmology of $\Omega_M=0.3$, $\Omega_\Lambda=0.7$, $h=0.7$; implying ${\rm DM}=40.44$ mag.)

While the spectrum can be reasonably approximated by a hot blackbody, some deviation from a simple blackbody curve is evident in Figure~\ref{fig:spectrum}.
There is some uncertainty in the exact shape of the spectrum due to uncertain atmospheric corrections and wavelength-dependent slit losses as well as limited wavelength overlap between the blue and red arms of the spectrograph.  However, both LRIS spectra and the close-to-simultaneous LT photometry self-consistently suggest a depression in flux in the vicinity of the observed $g$-band (rest-frame wavelengths between 3500-4500 \AA), similar to what was observed in spectra of AT2018cow at similar phases after peak \citep{Perley+2019}.  A spectrum of AT2018cow at a similar phase (from the Discovery Channel Telescope about six days after the peak; the flux has been rescaled) is shown for comparison in Figure~\ref{fig:spectrum}.

\begin{table}
	\centering
	\caption{Photometry of AT2020xnd. (The full table, including upper limits, is available online as supplementary material.)}
	\label{tab:photometry}
	\begin{tabular}{lllrrr}
		\hline
MJD & facility & filter & mag$^{a}$ & unc. & mag$^{b}$ \\
		\hline
59134.17188 & P48+ZTF    & r    &  20.52 & 0.14 &  20.34 \\
59134.18359 & P48+ZTF    & r    &  20.08 & 0.08 &  19.90 \\
59134.22656 & P48+ZTF    & g    &  19.68 & 0.06 &  19.43 \\
59134.22656 & P48+ZTF    & g    &  19.69 & 0.06 &  19.43 \\
59135.26953 & P48+ZTF    & r    &  19.74 & 0.07 &  19.56 \\
59136.17578 & P48+ZTF    & r    &  19.87 & 0.08 &  19.70 \\
59136.21094 & P48+ZTF    & g    &  19.49 & 0.05 &  19.23 \\
59136.21484 & P48+ZTF    & g    &  19.51 & 0.05 &  19.26 \\
59138.13281 & P48+ZTF    & r    &  19.92 & 0.09 &  19.74 \\
59138.19531 & P48+ZTF    & g    &  20.00 & 0.08 &  19.74 \\
59138.19531 & P48+ZTF    & g    &  20.07 & 0.09 &  19.81 \\
59139.00391 & LT+IOO     & g    &  20.22 & 0.02 &  19.97 \\
59139.00781 & LT+IOO     & r    &  20.28 & 0.04 &  20.11 \\
59139.00781 & LT+IOO     & i    &  20.47 & 0.05 &  20.34 \\
59139.00781 & LT+IOO     & z    &  20.45 & 0.13 &  20.36 \\
59139.01172 & LT+IOO     & u    &  19.93 & 0.05 &  19.59 \\
59139.19141 & P48+ZTF    & g    &  20.07 & 0.08 &  19.82 \\
59139.85938 & LT+IOO     & g    &  20.54 & 0.07 &  20.28 \\
59139.85938 & LT+IOO     & r    &  20.51 & 0.04 &  20.34 \\
59139.85938 & LT+IOO     & i    &  20.79 & 0.07 &  20.66 \\
59139.86328 & LT+IOO     & z    &  20.79 & 0.18 &  20.70 \\
59139.86328 & LT+IOO     & u    &  20.14 & 0.07 &  19.80 \\
59140.16797 & P48+ZTF    & g    &  20.66 & 0.11 &  20.41 \\
59140.17969 & P48+ZTF    & g    &  20.53 & 0.10 &  20.28 \\
59140.23047 & P48+ZTF    & r    &  20.78 & 0.14 &  20.60 \\
59141.17188 & P48+ZTF    & r    &  20.91 & 0.22 &  20.74 \\
59142.09375 & P60+SEDM   & r    &  21.01 & 0.13 &  20.84 \\
59142.09766 & P60+SEDM   & g    &  21.09 & 0.09 &  20.83 \\
59142.14453 & P48+ZTF    & r    &  21.01 & 0.16 &  20.83 \\
59142.14453 & P48+ZTF    & r    &  21.14 & 0.19 &  20.97 \\
59142.62891 & GIT        & r    &  21.63 & 0.06 &  21.46 \\
59143.10156 & P60+SEDM   & r    &  21.38 & 0.13 &  21.20 \\
59143.10547 & P60+SEDM   & g    &  21.68 & 0.12 &  21.42 \\
59143.10938 & P60+SEDM   & i    &  21.39 & 0.18 &  21.26 \\
59143.20703 & P48+ZTF    & g    &  21.31 & 0.19 &  21.06 \\
59143.22656 & P48+ZTF    & r    &  21.48 & 0.21 &  21.30 \\
59143.70703 & GIT        & r    &  21.82 & 0.06 &  21.65 \\
59144.41406 & Swift+UVOT & u    &  21.38 & 0.20 &  21.04 \\
59145.06641 & NTT+EFOSC2 & u    &  21.61 & 0.07 &  21.27 \\
59145.07422 & NTT+EFOSC2 & g    &  21.89 & 0.04 &  21.64 \\
59145.08203 & NTT+EFOSC2 & r    &  22.04 & 0.04 &  21.86 \\
59145.08984 & NTT+EFOSC2 & i    &  22.14 & 0.06 &  22.01 \\
59145.67578 & GIT        & r    &  22.31 & 0.08 &  22.13 \\
59147.04297 & NTT+EFOSC2 & r    &  22.64 & 0.12 &  22.46 \\
59147.05469 & NTT+EFOSC2 & i    &  22.99 & 0.18 &  22.86 \\
59147.06641 & NTT+EFOSC2 & g    &  22.37 & 0.07 &  22.12 \\
59151.46875 & Swift+UVOT & uvw1 &  23.19 & 0.28 &  22.73 \\
59156.03125 & VLT+FORS2  & g    &  23.60 & 0.03 &  23.35 \\
59156.04297 & VLT+FORS2  & R    &  23.65 & 0.07 &  23.49 \\
59156.05078 & VLT+FORS2  & I    &  23.69 & 0.10 &  23.57 \\
59157.02734 & VLT+FORS2  & g    &  23.82 & 0.04 &  23.57 \\
59157.03906 & VLT+FORS2  & R    &  23.90 & 0.07 &  23.74 \\
59157.05078 & VLT+FORS2  & I    &  24.25 & 0.13 &  24.13 \\
59158.06641 & VLT+FORS2  & g    &  23.99 & 0.04 &  23.74 \\
59158.07812 & VLT+FORS2  & R    &  23.96 & 0.06 &  23.80 \\
59158.08984 & VLT+FORS2  & I    &  23.94 & 0.11 &  23.82 \\
59162.06250 & VLT+FORS2  & g    &  24.27 & 0.06 &  24.02 \\
59162.07422 & VLT+FORS2  & R    &  24.30 & 0.09 &  24.14 \\
59162.08594 & VLT+FORS2  & I    &  24.23 & 0.15 &  24.11 \\
59167.02344 & VLT+FORS2  & u    &  24.95 & 0.20 &  24.61 \\
59167.04297 & VLT+FORS2  & g    &  24.78 & 0.11 &  24.53 \\
59167.04688 & VLT+FORS2  & R    &  24.86 & 0.16 &  24.70 \\
59167.05469 & VLT+FORS2  & I    &  24.97 & 0.25 &  24.85 \\
		\hline
\multicolumn{6}{l}{$^a$ AB magnitude, not corrected for extinction.}\\
\multicolumn{6}{l}{$^b$ AB magnitude, corrected for Galactic extinction.}\\
		\hline
	\end{tabular}
\end{table}

\subsection{Multiwavelength Observations}

After the confirmation of AT2020xnd as an extragalactic, fast-evolving transient we also obtained extensive observations at radio, millimetre, and X-ray wavelengths.
The transient is luminous at radio, millimeter, and X-ray wavelengths \citep{TNS204,TNS218}, further confirming its similarity to AT2018cow.  The inferred properties of the forward shock will be presented in separate work by Ho et al.
 
As part of our multiwavelength campaign, we obtained two sets of observations with the Neil Geherels Swift Observatory ({\it Swift}), using the X-Ray Telescope (XRT) and Ultraviolet-Optical Telescope (UVOT) simultaneously.  The first UVOT observation was conducted in $u$-band and the second observation in the $uvw1$-band.  The transient is only marginally detected in these exposures; the associated photometry is included in Table~\ref{tab:photometry}.

\section{Analysis and Discussion}

\subsection{A Fast-Peaking Optical Transient}

The optical light curve of 2020xnd is plotted in Figure~\ref{fig:lightcurve}.  Overplotted for reference as solid lines is the light curve of AT2018cow, taken from \cite{Perley+2019} and interpolated using the same procedure as in that work.  Colours are matched approximately by rest-frame filter bandpass (for AT2018cow we plot $uvw1$, $u$, $g$, and $r$ to match the colours of $u$, $g$, $r$, and $i$, respectively).  No offsets have been applied, although the reference time ($t_0$ at MJD 59132.0) was chosen for the best match to AT2018cow.  

While there is some uncertainty regarding the exact time of the peak and the nature of the rise (due to limited sampling and non-negligible photometric errors), the light curves are strikingly similar.  The peak absolute AB magnitude ($M_\lambda$) of AT\,2020xnd is approximately $M_{5000} = -20.6$ or $M_{3900} = -20.9$.  The total time above half peak is about 6 days as observed (two days to rise from half-peak, then four days to decay an equivalent amount) or 5 days in the rest-frame.  These values are quite similar to those inferred for AT2018cow ($M_{4600}=-20.4$, $t_{\rm 1/2} \sim 4.5$d; \citealt{Perley+2019}). No other classified ZTF transients, save for AT\,2018cow itself and the analog event ZTF\,18abvkwla \citep{Perley+2020,Ho+2020,Ho+2021}, share these properties.

\subsection{A Persistently Blue, Hot Transient}

The spectrum and spectral energy distribution (SED) of AT2020xnd imply a hot photosphere peaking well into the ultraviolet that persists throughout the observed evolution of the transient.  In Figure~\ref{fig:spectrum} we show both epochs of spectroscopy and SEDs from two (close in time) epochs of Liverpool Telescope photometry.  

We fit the multiband photometry from the first two LT epochs (at $7.01$ and $7.86$ observer-frame days after $t_0$) and the first NTT epoch (at $13.08$ days) to a simple blackbody model.  The effective temperature across all three epochs is very high ($T = 20000 \pm 2000$~K) and does not evolve between the first and last epochs to within the uncertainties.  The effective blackbody radius moves inward, from $R = 80 \pm 30$~AU at 7 days to $R = 39 \pm 5$~AU at 13 days.\footnote{Since the host extinction is unknown, these are technically lower limits on the temperature and upper limits on the radius.}

These are similar values to AT2018cow at equivalent times \citep{Perley+2019} and the physical implications are also similar: the photosphere is inconsistent with a dense sphere of expanding ejecta such as that seen in ordinary supernovae.  While a recessing photosphere can occur within an expanding explosion if the opacity drops quickly enough, it would be surprising for this to happen in material that remains far above the recombination temperatures of the elements likely to dominate the ejecta (H, He, C, and O) at small radii.  
We conclude that, as was the case for AT2018cow, the emission from AT2020xnd originates from a different component, such as a confined ejecta torus close to the central engine or dense shell of pre-existing material.

\subsection{No Evidence For A Radioactively-Powered Supernova}
\label{sec:snlimits}

The successful explosion of a massive star is expected to produce a large amount ($\gtrsim1$\,$M_\odot$) of hot, dense, freely-expanding ejecta.  In the previous section, we argue that such a component cannot reproduce the early-time evolution of AT2020xnd.  This on its own does not rule out the classical explosion model: the luminous, blue emission seen at early times could originate from shock-heated circumstellar material that simply outshines the more slowly-evolving light curve from the radioactively-heated ejecta underneath.  However, under this scenario we might still expect to see emission from the ejecta at later times once the hot blue component fades.

\begin{figure}
	\includegraphics[width=\columnwidth]{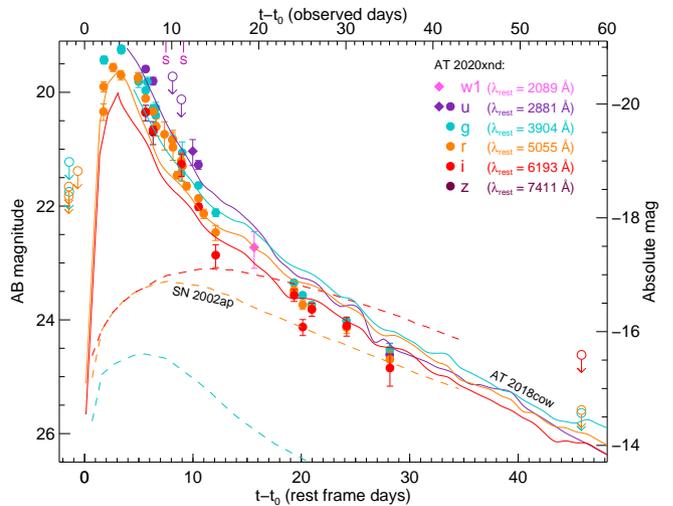}
    \caption{Optical/UV light curve of AT2020xnd (data points), compared with rest-frame equivalent light curves of AT2018cow (solid curves, based on the interpolation in \citealt{Perley+2019}) and the low-luminosity SN Ic-BL SN2002ap (dashed curve; from \citealt{Mazzali2002}). UVOT observations are shown as diamonds and ground-based observations as circles.  Measurements in the optical bands are colour-coded according to the nearest SDSS filter.  All measurements have been corrected for Galactic extinction.  The `S' marks denote the epochs of the two spectra shown in Figure \ref{fig:spectrum}.  }
    \label{fig:lightcurve}
\end{figure}

The maximum optical luminosity of any ``classical'' (radioactively-heated, $M_{\rm ej} \gtrsim 1 M_\odot$) supernova component in AT2020xnd is $M>-16.5$ ($\nu L_\nu \sim 10^{42}$ erg/s) at $t\sim20$ rest-frame days post-explosion or $M>-15.5$ ($\nu L_\nu \sim 4 \times 10^{41}$ erg/s) at $t\sim28$ rest-frame days.  (The magnitude limits are essentially the same in the observed $g$, $R$, and $I$ bands, which correspond to rest-frame wavelengths between 3200--7000 \AA). These limits are conservative, since the rapid fading and blue color observed during this period suggest the blue component continues to completely dominate the flux.
This is below the level of the underluminous Type Ic 2002ap \citep{Mazzali2002,Foley2003}, which is shown for reference in Figure~\ref{fig:lightcurve}), and comparable to low-luminosity Type II SNe.

This implies that, if the progenitor is a massive star, it either produced minimal radioactive nickel or expelled very little material. 
The $M_{R,{\rm peak}} - M{(^{56}{\rm Ni})}$ relation of \cite{Dessart+2016} would imply a limit of $M({^{56}{\rm Ni})} < 0.02 M_{\odot}$, given the measurements above and assuming that the Nickel-powered SN component peaked at around $\sim$20 days.  A higher nickel mass could be accommodated if the SN peaked significantly earlier, but this would in turn require a low ejecta mass: using the rise-time relation of \cite{Rest+2018} and scaling the light curve of SN\,2002ap \citep{Mazzali2002}, an ejecta mass of $M_{\rm ej} \lesssim 1 M_\odot$ would be needed, in tension with the notion of an exploding massive star.

Low-luminosity, nickel-poor supernovae are not unprecedented \citep{Hamuy2003,Pastorello+2004}, and nickel and ejecta mass estimates based on simple light curve properties are subject to large systematic uncertainties \citep{Khatami2019}.  The light curve properties of AT\,2020xnd alone thus cannot strictly rule out all known classes of massive stellar explosion on their own.  However, the results are striking in the context of the luminous radio emission (Ho et al. 2021, in prep), which probes the fastest ejecta.
Known radio-luminous transients associated with the deaths of massive stars are without exception accompanied by quite energetic supernovae \citep{Weiler2002}, with large ejecta and nickel masses.  In contrast, for both of the two radio-luminous fast blue transients with deep late-time follow-up reported so far, a classical supernova can be excluded to deep limits.

This suggests that events similar to AT2018cow and AT2020xnd may be fundamentally distinct from the types of explosions responsible for known radio-loud supernovae.  One possibility is that the progenitor is not a massive star at all, although this model is at odds with the dwarf star-forming nature of the host population (see next section).  Another possibility is that the fast radio shock and weak or absent SN are consequences of an explosion mechanism that is distinct from common classes of SNe, as in the black-hole collapse model.  More luminous, fast-evolving events will need to be studied in detail out to late times to firmly test this possibility.

\subsection{A Dwarf Host Galaxy With Modest Star-Formation}

While a sizeable population of photometrically-identified fast-luminous transients exists and has now been subject to detailed sample analysis \citep{Pursiainen+2018,Wiseman+2020}, it is not yet clear how many of these are true AT2018cow-like events versus other fast phenomena (such as classical Type Ibn SNe, or ``ordinary'' SN types with luminous shock-cooling peaks whose secondary rise was missed).
The sample of confirmed AT2018cow-like transients remains very small, although all three events published to date were localized to star-forming dwarf galaxies \citep{Perley+2020,Ho+2020,Coppejans+2020} and this appears to be true of the broader (spectroscopically-unconfirmed) fast-luminous transient population from DES as well \citep{Wiseman+2020}.

AT2020xnd continues this trend.  There is insufficient data to fit an SED model to the host photometry, although the optical luminosity (absolute magnitude $-16.2$) is consistent with a stellar mass between $3\times10^7 M_\odot$ and $3\times10^8 M_\odot$ \citep{Blanton+2011} and typical of galaxies with stellar mass of $\sim$ $10^8$ $M_\odot$, similar to the Small Magellanic Cloud.  The star-formation rate is low: assuming that the weak emission feature seen in our LRIS spectra is indeed H$\alpha$ emission from the galaxy (and assuming no host extinction), we measure a star-formation rate of SFR = $0.020 \pm 0.005 M_\odot$ yr$^{-1}$, which is fairly characteristic of SN host galaxies of this mass at low redshift \citep{Taggart+2021}.

\begin{figure}
	\includegraphics[width=\columnwidth]{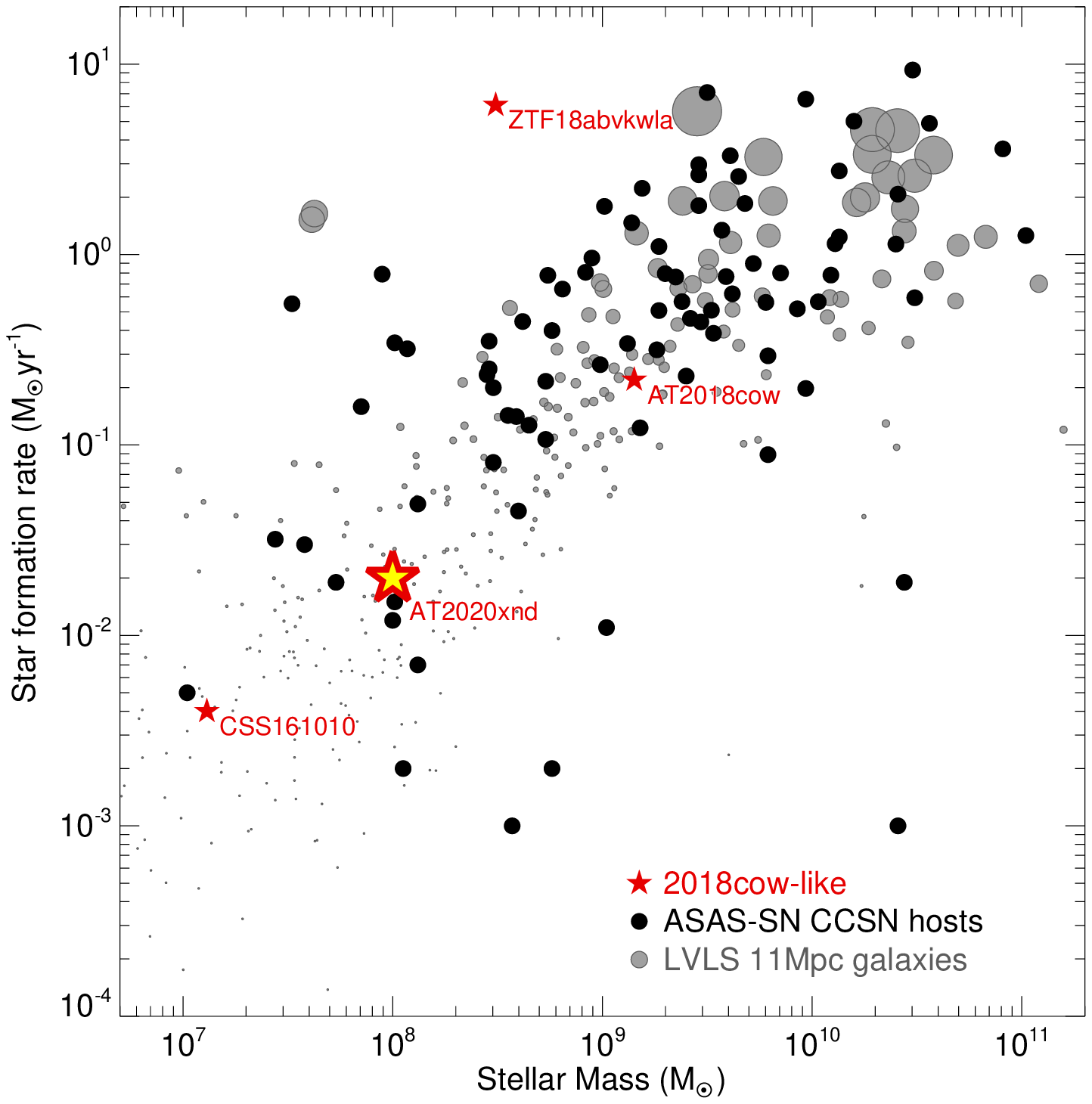}
    \caption{Estimates of the mass and star-formation rate of the host galaxy of AT2020xnd (star), compared to other hosts of AT2018cow-like events:  AT2018cow itself, ZTF18abvkwla, and CSS161010 \citep{Perley+2019,Ho+2019,Coppejans+2020}.  To date there are no high-mass hosts, and three out of four have typical star-formation rates given their stellar mass.  The core-collapse host galaxy sample of \citet{Taggart+2021} and galaxies within 11 Mpc from the Local Volume Legacy Survey (LVLS; \citealt{Lee+2011}) are also shown for comparison (in black and in grey, respectively).  LVLS symbol sizes are weighted by SFR for better visual comparison to the SFR-selected SN samples.}
    \label{fig:host}
\end{figure}

These observations continue to build the case that AT2018cow-like transients occur primarily (perhaps, exclusively) in dwarf galaxies, similar to superluminous supernovae and long-duration gamma-ray bursts.  The modest specific star formation rate suggests that an extremely young population age or high volumetric star-formation rate density is \emph{not} a precondition, which disfavours (although does not rule out) models that require a modified IMF or extensive dynamical interactions, effects expected only in the most extreme star-forming environments such as proto-globular clusters.  The basic physical properties of the host galaxies for the four AT2018cow-like events with confirmed luminous radio emission are shown in Figure~\ref{fig:host}.   While all four are low-mass galaxies, only ZTF18abvkwla has been shown to be forming stars at an elevated rate.   Recently-published IFU spectroscopy and millimetre observations of the host of AT2018cow \citep{Lyman+2020,Morokuma+2019} likewise argue against the notion of an elevated stellar or star-formation-rate density being essential for the production of the progenitor of this class.

\section{Conclusions}

We report the discovery and early characterization of AT 2020xnd, a fast-luminous optical transient at $z=0.2433$.  The transient shares all key properties with AT2018cow:  fast rise, high peak luminosity, featureless thermal spectrum, persistent blue colour throughout the decay, luminous radio emission, and a low-mass host galaxy whose star-formation rate is not particularly elevated.  Its spectrum may contain high-velocity photospheric features at early times, although we do not know whether it showed narrow emission components at late times.

These observations suggest that the peculiar properties of AT2018cow are in fact typical of a new class of fast, energetic transients.  They also reveal a sharp distinction between these events and others that occupy the same general region of luminosity-duration phase space \citep{Perley+2020}: for example Type Ibn supernovae and shock-cooling emission from Type IIb supernovae, whose optical properties are somewhat less extreme and do not produce luminous radio emission.  Understanding the nature of this mysterious class of AT2018cow-like astrophysical transients will require an integrative model that explains all of their key features: in particular, the strong contrast between an energetic high-velocity shock with the absence of an observable nickel-powered supernova.   

An updated discussion of progenitor models in the complete context of the optical, radio, and X-ray data of this event will be deferred for forthcoming work.  However, the evidence presented here points towards the failed-supernova scenario first proposed by \cite{Perley+2019}, in which the transient is produced by a jet driven by fallback accretion onto a black hole.  The association with low-mass but otherwise typical star-forming galaxies argues for a massive stellar origin (and against alternative models involving tidal disruption of a star around a pre-existing intermediate-mass black hole), and the absence of a classical SN counterpart implies that most of the star's mass was not ejected.  The hydrogen- and helium-rich late-time spectra of AT2018cow also support this model, although observations of this type could not be obtained for AT2020xnd. 

If this model is correct, we should continue to observe the key features of AT2018cow and AT2020xnd in further members of this newly-recognized class of events.   Our observations demonstrate an effective method for rapidly identifying new candidates in the future.  A rapid rise to flux well above that of its host galaxy, followed by immediate fading, is an identifying feature---although as this property is shared with cataclysmic variables additional criteria are necessary.  Angular extension of the candidate host in deep imaging surveys can provide an effective means of eliminating the majority of Galactic cataclysmic variables from consideration, with multi-band photometric follow-up with 1--2 meter class telescopes immediately after peak providing a useful additional screening method to rapidly vet candidates for spectroscopy.  

Even more robust filtering of false positives would be possible if reliable multi-colour host photometric redshifts, or even spectroscopic redshifts, of intermediate-redshift faint galaxies were widely available across most of the sky.  While the apparently high-redshift nature of its host galaxy (as given by the Legacy Survey) allowed AT2020xnd to pass our filter, this was in some ways a coincidence: the true redshift was lower than the (wide) constraint provided by the 3-band Legacy Survey.  Fortunately, over the coming decade the redshift completeness is poised to rapidly increase, with several large galaxy surveys in preparation including DESI, 4MOST, VLT/MOONS, and Rubin/LSST.  When combined with future high-cadence wide-field surveys similar to ZTF, this will allow far more robust redshift/luminosity-based filtering than is currently possible while also permitting candidates to be identified and followed up even more rapidly.

\section*{Acknowledgements}

We thank the referee for a constructive report that substantially improved the paper.
Based on observations obtained with the Samuel Oschin Telescope 48-inch and the 60-inch Telescope at the Palomar Observatory as part of the Zwicky Transient Facility project. ZTF is supported by the National Science Foundation under Grant No. AST-2034437 and a collaboration including Caltech, IPAC, the Weizmann Institute for Science, the Oskar Klein Centre at Stockholm University, the University of Maryland, Deutsches Elektronen-Synchrotron and Humboldt University, the TANGO Consortium of Taiwan, the University of Wisconsin at Milwaukee, Trinity College Dublin, Lawrence Livermore National Laboratories, and IN2P3, France. Operations are conducted by COO, IPAC, and UW.
The Liverpool Telescope is operated on the island of La Palma by Liverpool John Moores University in the Spanish Observatorio del Roque de los Muchachos of the Instituto de Astrofisica de Canarias with financial support from the UK Science and Technology Facilities Council. 
The SED Machine is based upon work supported by the National Science Foundation under Grant No. 1106171.
Based on observations collected at the European Organisation for Astronomical Research in the Southern Hemisphere under ESO programme(s) 106.21U2 and 106.216C.  We thank Stephane Blondin for helpful comments.

This work was supported by the GROWTH (Global Relay of Observatories Watching Transients Happen) project funded by the NSF under PIRE grant 1545949. GROWTH is a collaborative project among California Institute of Technology (USA), University of Maryland College Park (USA), University of Wisconsin Milwaukee (USA), Texas Tech University (USA), San Diego State University (USA), University of Washington (USA), Los Alamos National Laboratory (USA), Tokyo Institute of Technology (Japan), National Central University (Taiwan), Indian Institute of Astrophysics (India), Indian Institute of Technology Bombay (India), Weizmann Institute of Science (Israel), The Oskar Klein Centre at Stockholm University (Sweden), Humboldt University (Germany), Liverpool John Moores University (UK), and University of Sydney (Australia).

The GROWTH India telescope is a 70-cm telescope with a 0.7 degree field of view, set up by the Indian Institute of Astrophysics and the Indian Institute of Technology Bombay with support from the Indo-US Science and Technology Forum (IUSSTF) and the Science and Engineering Research Board (SERB) of the Department of Science and Technology (DST), Government of India. It is located at the Indian Astronomical Observatory (Hanle), operated by the Indian Institute of Astrophysics (IIA). GROWTH-India project is supported by SERB and administered by IUSSTF.

Funding for the SDSS and SDSS-II has been provided by the Alfred P. Sloan Foundation, the Participating Institutions, the NSF, the U.S. Department of Energy, the National Aeronautics and Space Administration (NASA), the Japanese Monbukagakusho, the Max Planck Society, and the Higher Education Funding Council for England.     The SDSS is managed by the Astrophysical Research Consortium for the Participating Institutions. The Participating Institutions are the American Museum of Natural History, Astrophysical Institute Potsdam, University of Basel, University of Cambridge, Case Western Reserve University, University of Chicago, Drexel University, Fermilab, the Institute for Advanced Study, the Japan Participation Group, Johns Hopkins University, the Joint Institute for Nuclear Astrophysics, the Kavli Institute for Particle Astrophysics and Cosmology, the Korean Scientist Group, the Chinese Academy of Sciences (LAMOST), Los Alamos National Laboratory, the Max-Planck-Institute for Astronomy (MPIA), the Max-Planck-Institute for Astrophysics (MPA), New Mexico State University, Ohio State University, University of Pittsburgh, University of Portsmouth, Princeton University, the United States Naval Observatory, and the University of Washington.

The Pan-STARRS1 Surveys (PS1) have been made possible through contributions of the Institute for Astronomy, the University of Hawaii, the Pan-STARRS Project Office, the Max-Planck Society and its participating institutes, the Max Planck Institute for Astronomy, Heidelberg and the Max Planck Institute for Extraterrestrial Physics, Garching, The Johns Hopkins University, Durham University, the University of Edinburgh, Queen's University Belfast, the Harvard-Smithsonian Center for Astrophysics, the Las Cumbres Observatory Global Telescope Network Incorporated, the National Central University of Taiwan, the Space Telescope Science Institute, NASA under grant NNX08AR22G issued through the Planetary Science Division of the NASA Science Mission Directorate, the NSF under grant  AST-1238877, the University of Maryland, and Eotvos Lorand University (ELTE).

The Legacy Surveys consist of three individual and complementary projects: the Dark Energy Camera Legacy Survey (DECaLS; NSF's OIR Lab Proposal ID 2014B-0404; PIs: David Schlegel and Arjun Dey), the Beijing-Arizona Sky Survey (BASS; NSF's OIR Lab Proposal ID 2015A-0801; PIs: Zhou Xu and Xiaohui Fan), and the Mayall z-band Legacy Survey (MzLS; NSF's OIR Lab Proposal ID 2016A-0453; PI: Arjun Dey). DECaLS, BASS and MzLS together include data obtained, respectively, at the Blanco telescope, Cerro Tololo Inter-American Observatory, The NSF's National Optical-Infrared Astronomy Research Laboratory (NSF's OIR Lab); the Bok telescope, Steward Observatory, University of Arizona; and the Mayall telescope, Kitt Peak National Observatory, NSF's OIR Lab. The Legacy Surveys project is honored to be permitted to conduct astronomical research on Iolkam Du'ag (Kitt Peak), a mountain with particular significance to the Tohono O'odham Nation.

This project used data obtained with the Dark Energy Camera (DECam), which was constructed by the Dark Energy Survey (DES) collaboration. Funding for the DES Projects has been provided by the U.S. Department of Energy, the U.S. National Science Foundation, the Ministry of Science and Education of Spain, the Science and Technology Facilities Council of the United Kingdom, the Higher Education Funding Council for England, the National Center for Supercomputing Applications at the University of Illinois at Urbana-Champaign, the Kavli Institute of Cosmological Physics at the University of Chicago, Center for Cosmology and Astro-Particle Physics at the Ohio State University, the Mitchell Institute for Fundamental Physics and Astronomy at Texas A\&M University, Financiadora de Estudos e Projetos, Fundacao Carlos Chagas Filho de Amparo, Financiadora de Estudos e Projetos, Fundacao Carlos Chagas Filho de Amparo a Pesquisa do Estado do Rio de Janeiro, Conselho Nacional de Desenvolvimento Cientifico e Tecnologico and the Ministerio da Ciencia, Tecnologia e Inovacao, the Deutsche Forschungsgemeinschaft and the Collaborating Institutions in the Dark Energy Survey. The Collaborating Institutions are Argonne National Laboratory, the University of California at Santa Cruz, the University of Cambridge, Centro de Investigaciones Energeticas, Medioambientales y Tecnologicas-Madrid, the University of Chicago, University College London, the DES-Brazil Consortium, the University of Edinburgh, the Eidgenossische Technische Hochschule (ETH) Zurich, Fermi National Accelerator Laboratory, the University of Illinois at Urbana-Champaign, the Institut de Ciencies de l'Espai (IEEC/CSIC), the Institut de Fisica d'Altes Energies, Lawrence Berkeley National Laboratory, the Ludwig-Maximilians Universitat Munchen and the associated Excellence Cluster Universe, the University of Michigan, the National Optical Astronomy Observatory, the University of Nottingham, the Ohio State University, the University of Pennsylvania, the University of Portsmouth, SLAC National Accelerator Laboratory, Stanford University, the University of Sussex, and Texas A\&M University.

The Legacy Surveys imaging of the DESI footprint is supported by the Director, Office of Science, Office of High Energy Physics of the U.S. Department of Energy under Contract No. DE-AC02-05CH1123, by the National Energy Research Scientific Computing Center, a DOE Office of Science User Facility under the same contract; and by the U.S. National Science Foundation, Division of Astronomical Sciences under Contract No. AST-0950945 to NOAO.

L.G. was funded by the European Union's Horizon 2020 research and innovation programme under the Marie Sk\l{}odowska-Curie grant agreement No. 839090. This work has been partially supported by the Spanish grant PGC2018-095317-B-C21 within the European Funds for Regional Development (FEDER).  L.G. acknowledges financial support from the Spanish Ministry of Science, Innovation and Universities (MICIU) under the 2019 Ram\'on y Cajal program RYC2019-027683 and from the Spanish MICIU project PID2020-115253GA-I00.
H.K. thanks the LSSTC Data Science Fellowship Program, which is funded by LSSTC, NSF Cybertraining Grant \#1829740, the Brinson Foundation, and the Moore Foundation; his participation in the program has benefited this work.
E.C.K. acknowledges support from the G.R.E.A.T research environment funded by {\em Vetenskapsr\aa det}, the Swedish Research Council, under project number 2016-06012, and support from The Wenner-Gren Foundations.  M.N. is supported by a Royal Astronomical Society Research Fellowship.  M.G. is supported by the Polish NCN MAESTRO grant 2014/14/A/ST9/00121.  T.-W.C. acknowledges the EU Funding under Marie Sk\l{}odowska-Curie grant H2020-MSCA-IF-2018-842471.

This research has made use of the NASA/IPAC Extragalactic Database (NED),
which is operated by the Jet Propulsion Laboratory, California Institute of Technology, under contract with NASA.  This research has made use of the VizieR catalogue access tool, CDS, Strasbourg, France (DOI : 10.26093/cds/vizier). The original description of the VizieR service was published in A$\&$AS 143, 23.

Some of the data that contributed to this paper were obtained at the W. M. Keck Observatory, which is operated as a scientific partnership among the California Institute of Technology, the University of California, and NASA. The Observatory was made possible by the generous financial support of the W. M. Keck Foundation. The authors wish to recognize and acknowledge the very significant cultural role and reverence that the summit of Maunakea has always had within the indigenous Hawaiian community.

%%%%%%%%%%%%%%%%%%%%%%%%%%%%%%%%%%%%%%%%%%%%%%%%%%
\section*{Data Availability}

Photometry is provided in Table 1.   Reduced LRIS spectra are available on WISErep.  All P48, Liverpool Telescope, Swift, Keck, NTT, and VLT observations will be available online via their respective telescope archives. 


%\bibliographystyle{mnras}
%\bibliography{ref}
\include{ref}

%%%%%%%%%%%%%%%%%%%%%%%%%%%%%%%%%%%%%%%%%%%%%%%%%%

% Don't change these lines
\bsp	% typesetting comment
\label{lastpage}
\end{document}